\documentclass[namedreferences]{solarphysics}
\usepackage[hyperref,optionalrh,solaromanenum]{spr-sola-addons} 
\usepackage{graphicx}                    
\usepackage{amssymb}                    
\usepackage{color}                       
\usepackage{breakurl}                         

\hypersetup{
    colorlinks=true,
    linkcolor=blue,
    filecolor=magenta,   
    citecolor=blue,
    urlcolor=blue,
}

\begin{document}

\begin{article}

\begin{opening}

\title{A Statistical Study of Low-frequency Solar Radio Type-III Bursts}

%

%
\runningauthor{Mahender et al}
\runningtitle{A Statistical Study of Solar Radio Type-III Bursts}

\author[addressref={aff1},corref,email={mahenderastro@gmail.com}]{\inits{Mahender}\fnm{Mahender}~\lnm{Aroori}}
\author[addressref={aff2}, email={sasikumar.raja@iiap.res.in}]{\inits{K. Sasikumar Raja}\fnm{K. Sasikumar Raja}~\lnm{}}
\author[addressref=aff2,email={ramesh@iiap.res.in}]{\inits{R. Ramesh}\fnm{R. Ramesh}~\lnm{}}
\author[addressref=aff2,email={vemareddy@iiap.res.in}]{\inits{Vemareddy}\fnm{Vemareddy}~\lnm{Panditi}}

\author[addressref={aff3},email={monstein@irsol.ch}]{\inits{Christian Monstein}\fnm{Christian Monstein}~\lnm{}}

\author[addressref=aff1,email={gyh042000@yahoo.co.in}]{\inits{Yellaiah}\fnm{Yellaiah}~\lnm{Ganji}}

\address[id=aff1]{Department of Astronomy, Osmania University, Hyderabad, Telangana, 500 007, India.}
\address[id=aff2]{Indian Institute of Astrophysics, 2nd Block, Koramangala, Bangalore - 560 034, India.}
\address[id=aff3]{Istituto Ricerche Solari Locarno (IRSOL), Via Patocchi - Prato Pernice, 6605 Locarno Monti, Switzerland.}

\begin{abstract}
We have studied low-frequency (45 - 410 MHz) type III solar radio bursts observed using e-CALLISTO spectrometer located at Gauribidanur radio observatory, India during 2013 - 2017. After inspecting the 1531 type III bursts we found that 426 bursts were associated with flares, while the other bursts might have triggered by small scale features / weak energy events present in the solar corona. In this study, we have carried out a statistical analysis of various observational parameters like start time, lower and upper frequency cut-offs of type III bursts and their association with flares, variation of such parameters with flare parameters such as location, class, onset and peak timings. From this study, we found that most of the high frequency bursts (whose upper frequency cut-off $> 350$ MHz) are originated from the western longitudes. We interpret that it could be due to the facts that Parker spirals from these longitudes are directed towards the earth and high frequency bursts are more directive. Further we report that number of bursts that reach earth from western longitudes are higher than eastern longitudes.

\end{abstract}

%

\end{opening}

\section{Introduction} \label{Intro}

Among various types of solar radio bursts that are classified by \citet{Wild1950}, type III bursts are the most intense, fast drifting and frequently observed bursts. These bursts are observed ranging from inner corona to 1 AU and some times even beyond. Type III bursts occur as a isolated bursts (lasts from 1 to 3 s), groups that lasts in 10 minutes, and as a storm which lasts from 10 mins to few tens of hours. Although their emission mechanism is still debated, their occurrence happens in the two steps: (i) during a flare or magnetic re-connection, huge amount of magnetic energy converts into kinetic energy that results in acceleration of particles. Such particles generate plasma oscillation (also known as Langmuir waves) during their passage along open magnetic field lines in the solar corona and interplanetary medium (IPM); (ii) subsequent conversion of these oscillations into electro-magnetic waves at the plasma frequency ($f_p$) called fundamental (F) emission and harmonic (H) emission at $2f_p$ \citep[see for example][]{Ginzburg1958,Zhe1970,Mel1980,Ramesh2003,Ramesh2005,Rei2014,Kishore2015,Kishore2017}. The measured average frequency ratio of harmonic emission to fundamental emission is 1.8 \citep{Ste1974}. It was also reported that F emissions are more directive than the H emissions \citep{Suz1982, Sas2013}. The electron density ($N_e$) and hence plasma frequency ($f_p \propto \sqrt{N_e}$) decreases as we move radially outwords in the solar corona. Therefore, lower the frequency of observations we probe outer layers in the corona and hence, type III bursts drift from higher to lower frequencies.

In this article, we have carried out a statistical study of the 1531 type III bursts that are observed during 2013-2017. For this study, we have used the data obtained using the Compact Astronomical Low-frequency Low-cost Instrument for Spectroscopy in Transportable Observatories (CALLISTO) spectrometer\footnote{http://www.e-callisto.org/} \citep{Benz2009, Mon2007, Zucca2012, Sas2018} located at Gauribidanur Radio Observatory (GRO; Lat: $13^{\circ} 36' 12''$ N and Long: $77^{\circ} 27' 07''$ E) which is 100 km north of Bangalore, India \citep{Ramesh2011, Ramesh2014}. An extensive statistical study of the type III bursts has not been done so far because of lack of data sets and due to non-existence of the automated detection algorithms. 
In this work, we have make use of the catalog provided by \citet{Singh2019} that are detected using an automated algorithm. Earlier studies of type III bursts have been done by \citet{Saint2013} using the observations carried out using Nancay Radioheliograph observations \citep{Ker1997}. 

\section{Observations} 
\label{Obs}

In this study, we have used data observed by the CALLISTO spectrometer that is located at Gauribidanur Radio Observatory (GRO). A single log-periodic dipole antenna that operate over 30 - 1100 MHz (VSWR $\lesssim$ 2) with a gain of $\approx 8$ dB was used as a primary receiving element which is then fed to an amplifier of 45 dB at the base of the antenna. The signal brought to the receiver room which is about a 100 meter using coaxial cable and then connected to the receiver developed at ETH Zurich, Switzerland. Although the receiver can operate from 45 - 870 MHz, the CALLISTO spectrometer at GRO is configured to operate between 45 - 410 MHz in order to increase the frequency resolution. The frequency resolution of the spectrometer is 62.5 kHz and the radiometric bandwidth is $\approx 300$ kHz. The CALLISTO spectrometer was set up in 2009 and since then it has been used to monitor transient emissions from the solar corona during 02:30 - 11:30 UT. The time resolution of the instrument is 0.25s at the rate of 200 channels per spectrum. Note that the time difference between two consecutive spectral pixels is 1.25ms \citep{Benz2009, Mon2007, Singh2019}.\\

\begin{figure}[ht!]
\centering
\includegraphics[width=1.0\textwidth,clip=]{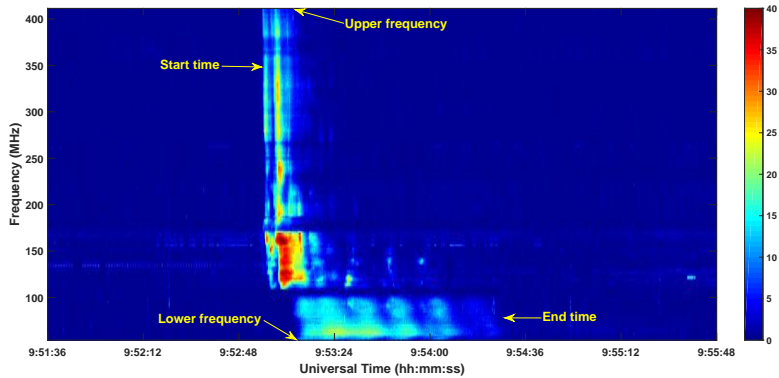}
\caption{The dynamic spectra of type-III burst observed on September 7, 2017 by the CALLISTO spectrometer located at GRO. The median of every frequency channel is subtracted accordingly. The arrows refer to the start and end time (in UT) and lower and upper frequencies (in MHz) that are measured manually in this study. Note that 88-108 MHz and around 170 MHz, band stop filters have been used to avoid RFI from the FM and TV stations.}

\label{fig:ds}
\end{figure}

\begin{figure}[ht!]
\includegraphics[width=0.9\textwidth,clip=]{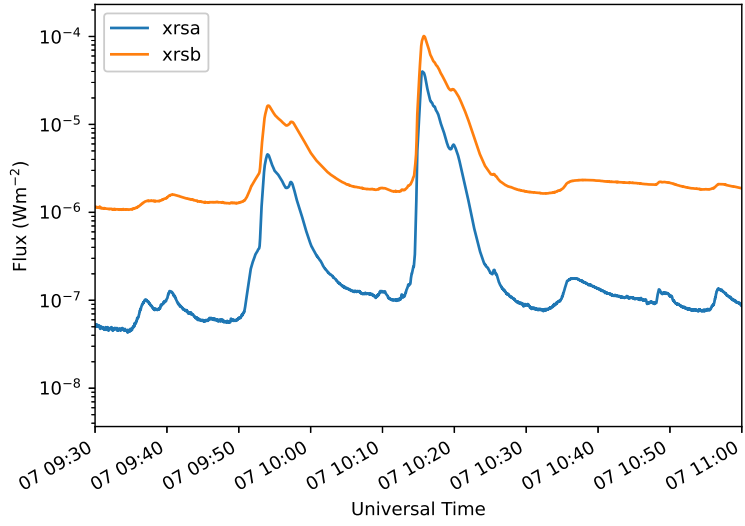}
\caption{The flare of class M1.4 observed by GOES on September 7, 2017 during 09:49 - 09:58 UT (with peak at 09:54 UT) that triggered type III burst shown in Figure \ref{fig:ds}.}

\label{fig:goes}
\end{figure}

\begin{figure}[ht!]
\includegraphics[width=1\textwidth,clip=]{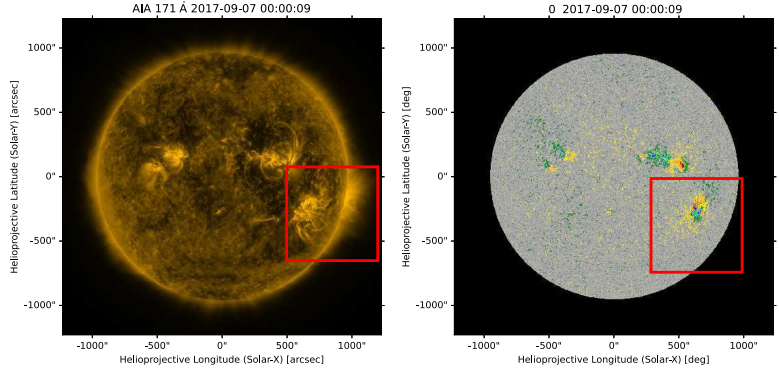}
\caption{The active region NOAA 12673 associated with the flare and type III burst during 09:49 - 09:58 UT. In the left panel, the full disk of the Sun observed at AIA 171~\AA~and in right panel observation of HMI magnetogram is shown. The red color box in both panels refer to the AR 12673 and its location is S07$^{o}$W46$^{o}$.}
\label{fig:sdo}
\end{figure}

In this study, we have used the catalog of type III bursts that was identified using automated algorithm by \citet{Singh2019} during 2013 - 2017 (\url{http://www.iiserpune.ac.in/~p.subramanian/Bursts.zip
}). For example, Figure~\ref{fig:ds} shows the dynamic spectrogram that was observed on September 7, 2017 using the CALLISTO spectrometer located at GRO. Note that median of the every frequency channel measured over a whole day was subtracted from the corresponding channel to get rid of continuous radio frequency interference (RFI). While preparing the catalog, in order to mitigate type III burst-like RFI, \citet{Singh2019} have eliminated the features / RFI with the drifting rates ($v_d= {{\Delta f} \over {\Delta t}}$, where $\Delta f$ and $\Delta t$ are bandwidth and duration of the burst/feature in the dynamic spectrogram) that does not fall in the range $0.81 \:{\rm MHz~s^{-1}} < v_d < 162\: \rm MHz~s^{-1}$. This filtering technique successfully eliminates the type III burst-like RFI without eliminating the type III bursts. In addition, we have measured start and end time of the type III burst shown in Figure \ref{fig:ds} which are 9:52:58 and 9:54:28 UT; and the lower and upper frequency cutoffs of burst are 45 to $> 410$ MHz. Similarly, we have identified 1531 type III bursts and measured these parameters manually. Furthermore, we have studied the flare associated type III bursts by knowing the flare parameters like onset, peak, end times, class of the flare, associated active region, and location of the flare \footnote{\url{https://cdaw.gsfc.nasa.gov/CME\_list/NOAA/org_events_text/}}$^{,}$ \footnote{\url{https://www.ngdc.noaa.gov/stp/space-weather/solar-data/solar-features/solar-flares/x-rays/goes/xrs/}}$^{,}$ \footnote{\url{http://www.lmsal.com/solarsoft/ssw/last_events-2014/}}$^{,}$ \footnote{\url{http://hec.helio-vo.eu/hec/hec_gui.php}}. 
For instance the type III bursts shown in Figure \ref{fig:ds} was triggered by a M1.4 class flare as observed by Geostationary Operational Environmental Satellites (GOES) as seen in Figure \ref{fig:goes}. The onset and end time of the flare were 09:49 and 09:58 UT with peak emission at 09:54 UT. In the Figure \ref{fig:sdo}, left panel shows the observations of Atmospheric Imaging Assembly \citep[AIA;][]{Lem2012} onboard  \textit{Solar Dynamics Observatory} \citep[SDO;][]{Pes2012} at 171~\AA  wavelength band and the right panel shows the observations of Helioseismic and Magnetic Imager \citep[HMI;][]{Sch2012} onboard SDO. The red colored boxes in both panels indicate the active region NOAA 12673 that triggered the type III burst (see Figure \ref{fig:ds}) and the flare (see Figure~\ref{fig:goes}). The location of the flare was at S07$^{o}$W46$^{o}$ (see Figure \ref{fig:sdo}).

Similarly, we have studied 1531 type III bursts and found that 426 of them were flare associated. By knowing the class, location of the flare, we have studied the way these parameters affects the lower and upper frequency cut-off of type III bursts. 

\section{Results and discussions}
\label{Res}

After a careful manual inspection of the catalog of type III bursts that are listed at \url{http://www.iiserpune.ac.in/~p.subramanian/Bursts.zip}, we have identified 1531 intense and well separated type III bursts observed in the frequency range 45 - 410 MHz during 2013- 2017. We have found that among them only 426 (28\%) bursts are associated with the flares. The remaining 1105 type III bursts (72\%) presumably originated due to small scale features that are present in solar corona and they are the signatures of weak energy releases reported earlier in literature \citep{Ram2010, Ram2013, James2017, James2018, Rohit2018, Mugundhan2016, Mugundhan2018, Singh2019}. 

\begin{figure}[ht!]
\includegraphics[width=0.9\textwidth,clip=]{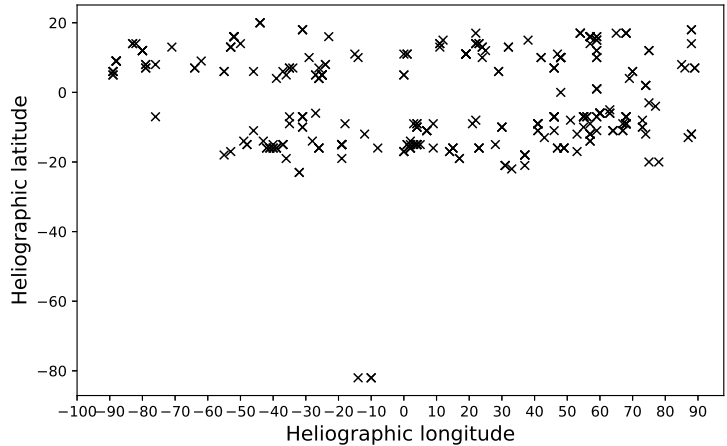}

\caption{The heliographic longitude  vs the heliographic latitude of flares that triggered type III bursts.}
\label{fig:longVsLat}
\end{figure}

We have then identified the type III bursts that are observed during onset - end time of the flares. Assuming that such bursts are associated with the flaring processes, we have measured the flare class, location and onset and end timings of the flares and studied the way these parameters vary with lower, upper frequency cutoff of type III bursts. Firstly, we have plotted the heliographic longitudes and latitudes of the flares associated with the type III bursts as shown in Figure \ref{fig:longVsLat}. We have found that most of the type III bursts originated close to the equator (i.e. heliographic latitudes $\pm 23^o$) except the three instances that occurred near the polar cap of the southern hemisphere.

\begin{figure*}
   \centerline{\hspace*{0.015\textwidth}
               \includegraphics[width=0.5\textwidth,clip=]{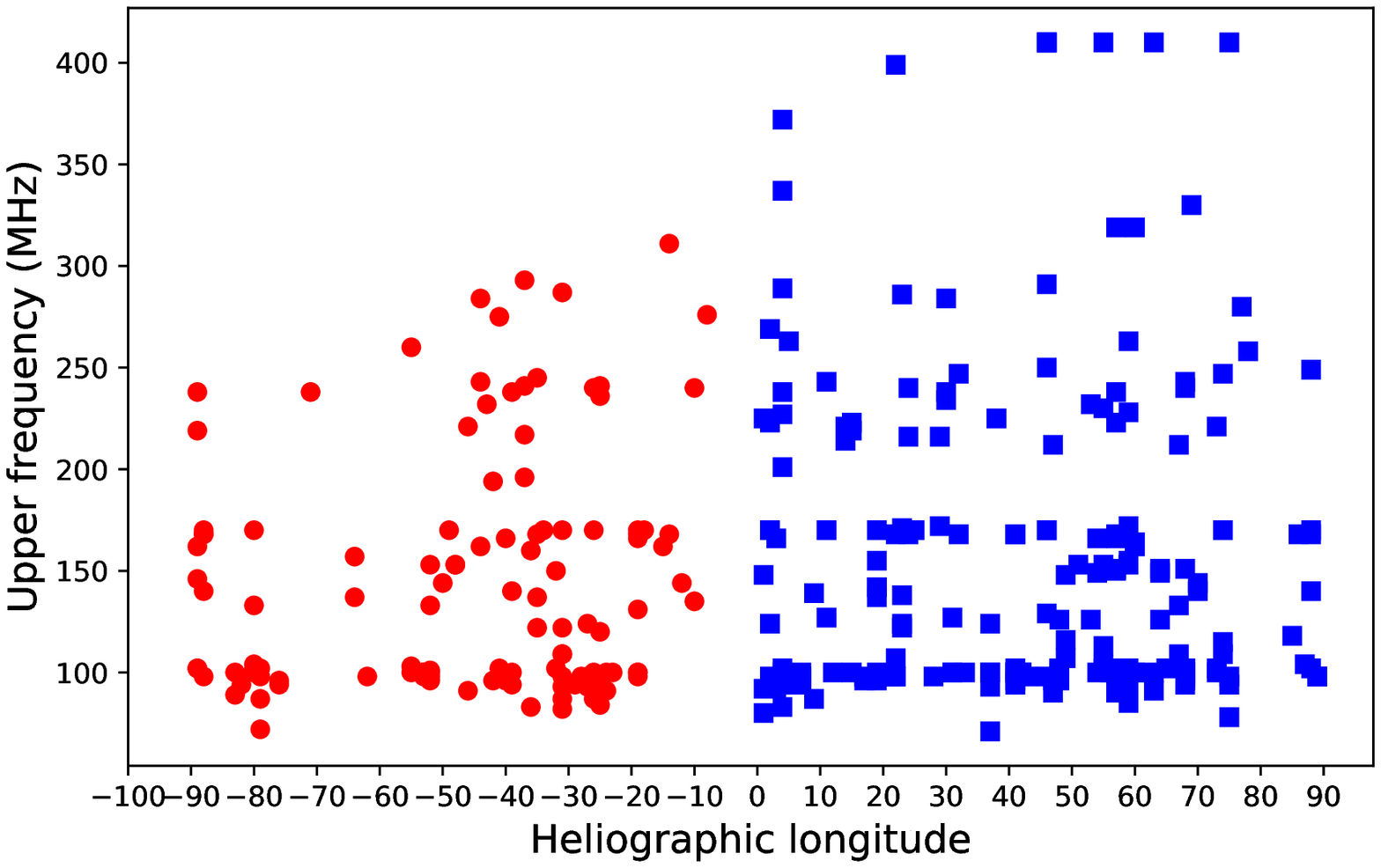}
               \hspace*{-0.06\textwidth}
               \includegraphics[width=0.5\textwidth,clip=]{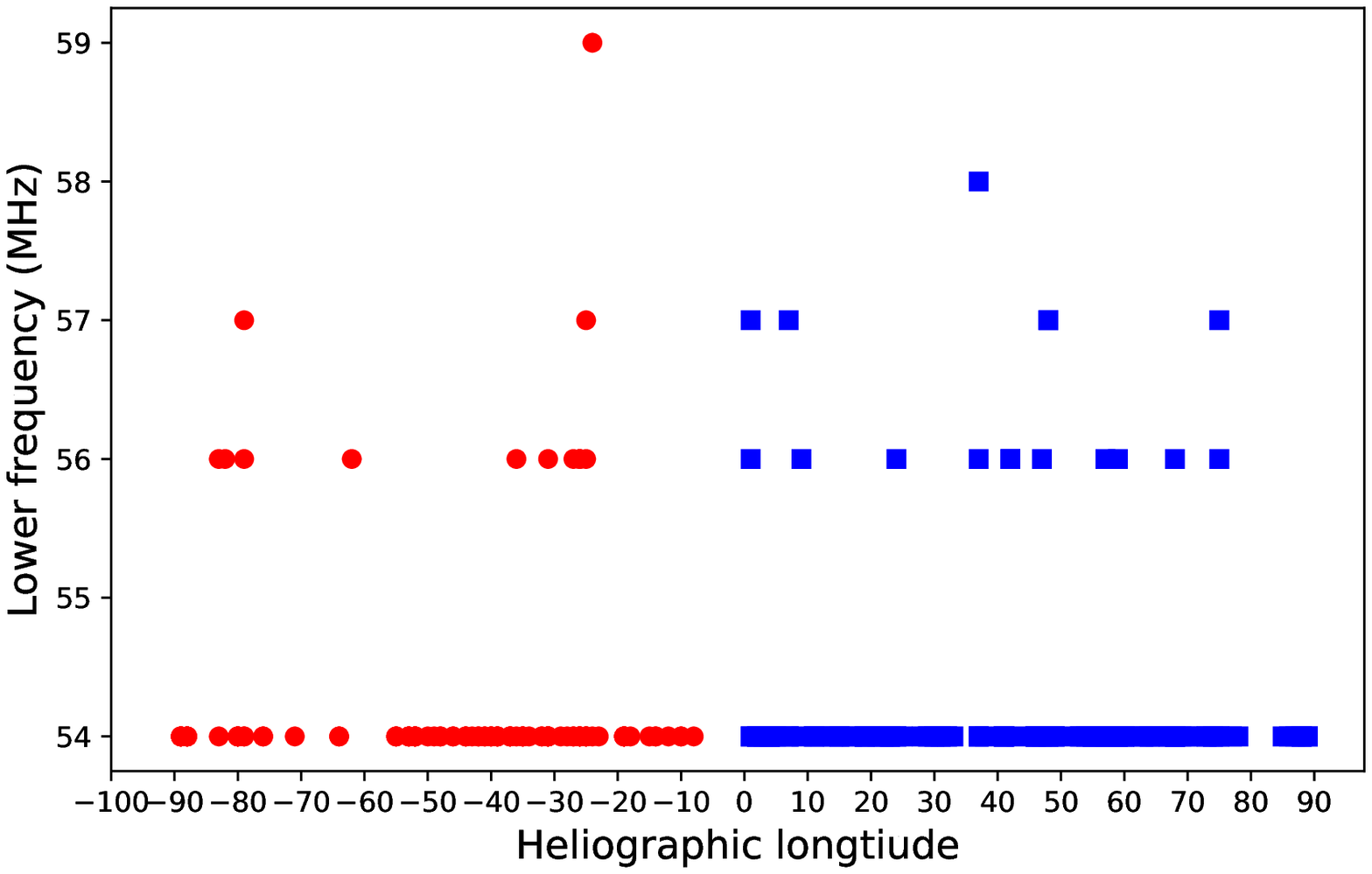}
              }
     \vspace{-0.26\textwidth}   
     \centerline{\Large \bf     
      \hspace{0.1 \textwidth}  \color{black}{(a)}
      \hspace{0.39\textwidth}  \color{black}{(b)}
         \hfill}
     \vspace{0.22\textwidth}    
   \centerline{\hspace*{0.015\textwidth}
               \includegraphics[width=0.5\textwidth,clip=]{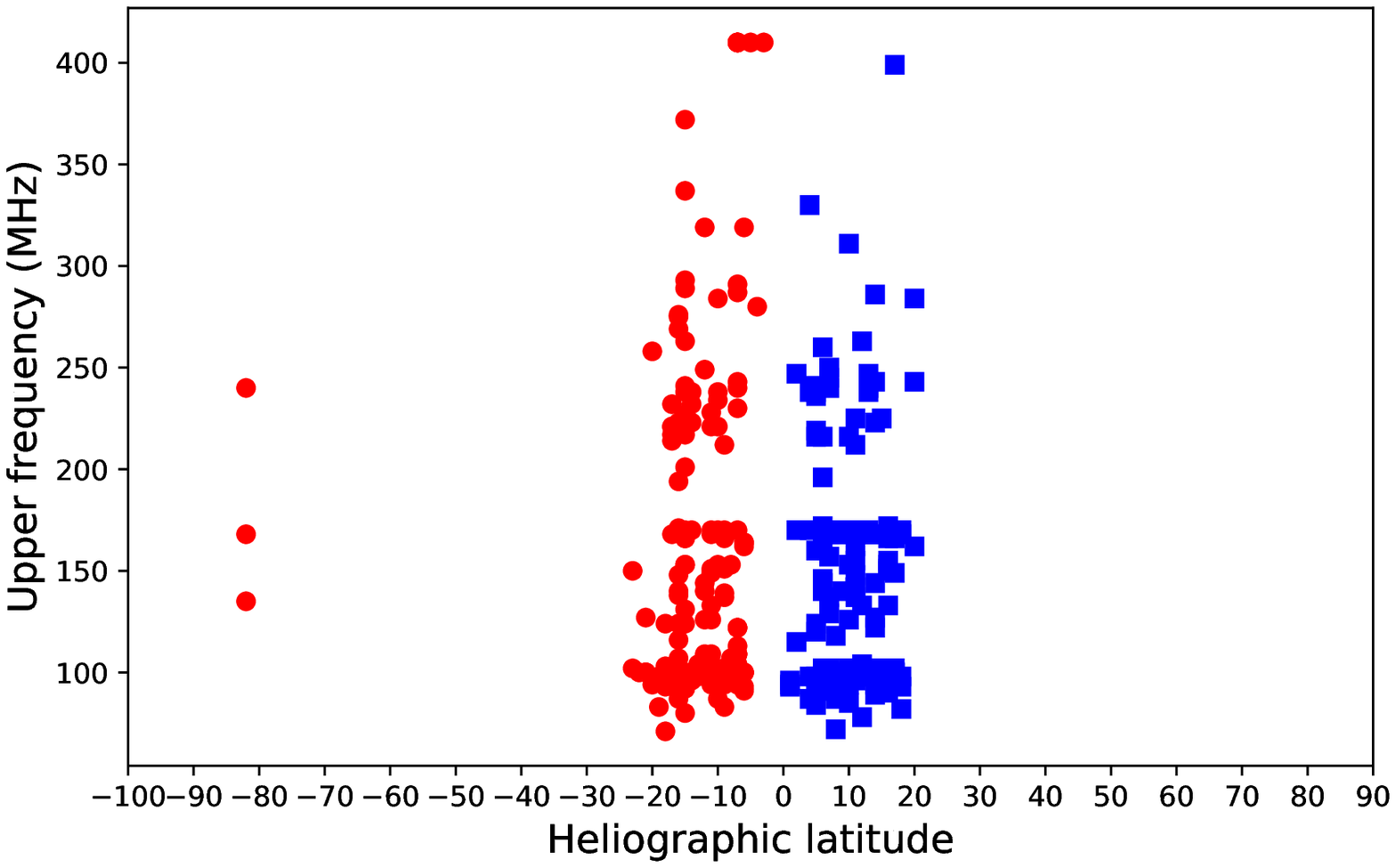}
               \hspace*{-0.06\textwidth}
               \includegraphics[width=0.5\textwidth,clip=]{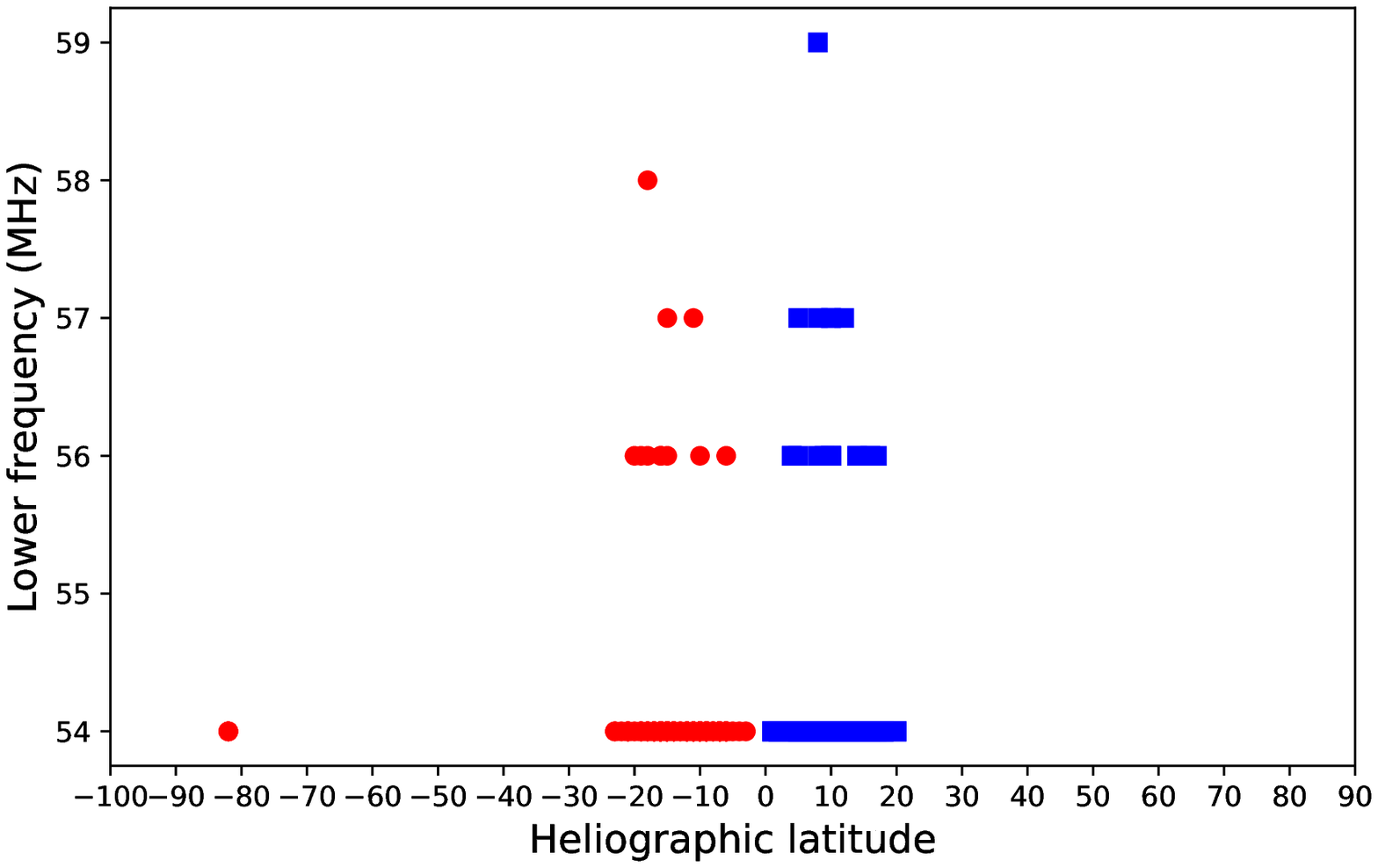}
              }
     \vspace{-0.26\textwidth}   
     \centerline{\Large \bf     
      \hspace{0.1 \textwidth} \color{black}{(c)}
      \hspace{0.39\textwidth}  \color{black}{(d)}
         \hfill}
     \vspace{0.22\textwidth}    
                
\caption{The Figure shows relationship between upper and lower frequency of type III bursts and their variation with heliographic longitude and latitude.  Panel (a) shows variation of upper frequency cut-off of radio burst with the heliographic longitude and panel (b) shows variation of lower frequency cut-off of type III bursts with the heliographic longitude. Similarly panel (c) shows variation of upper frequency cut-off with heliographic latitude and panel (d) shows variation of lower frequency cut-off of type III bursts with the heliographic latitude.}
\label{fig:long_freq}
\end{figure*}

Figure \ref{fig:long_freq} shows the variation of lower and upper frequency of the type III bursts with heliographic longitude and latitude of the associated flares. Panel (a) shows the variation of heliographic longitude of flare and its upper frequency of the burst. The red and blue colors indicate the east and west heliographic longitudes. It is clearly seen that type III bursts whose upper frequency cutoff is larger are originated from the west longitudes than the east longitudes. We interpret that, it is due to the following facts (i) the Parker spirals from the west longitudes are directed toward the earth, (ii) type III bursts possesses the directivity, and (iii) type III bursts are more directive at high frequencies than the low frequencies.  Panel (b) shows variation of lower frequency with the heliographic longitudes. It doesn't show any trend because of instrumental limitation (i.e. we can not observe below 45 MHz). Panels (c) and (d) shows variation of upper and lower frequencies with the heliographic latitudes respectively and they show that most of the flare associated type III bursts are originated in the heliographic latitude range of $\pm 23^\circ$. This study suggests also, that a few of the bursts are observed from southern polar cap regions. The red and blue color regions in panel (c) and (d) indicate the southern and northern latitudes respectively.

\begin{figure}[ht!]
\includegraphics[width=6cm,keepaspectratio]{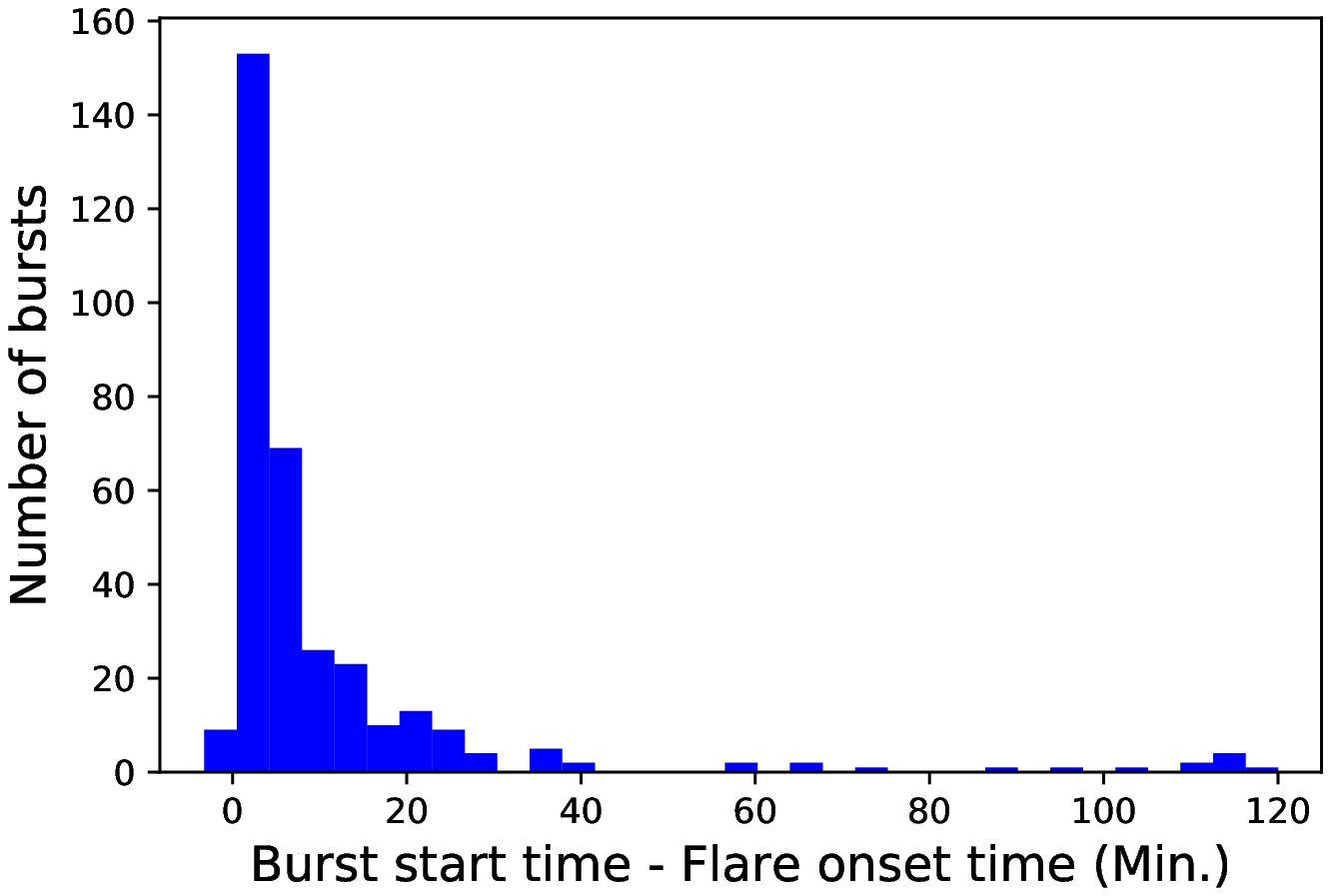}
\includegraphics[width=6cm,keepaspectratio]{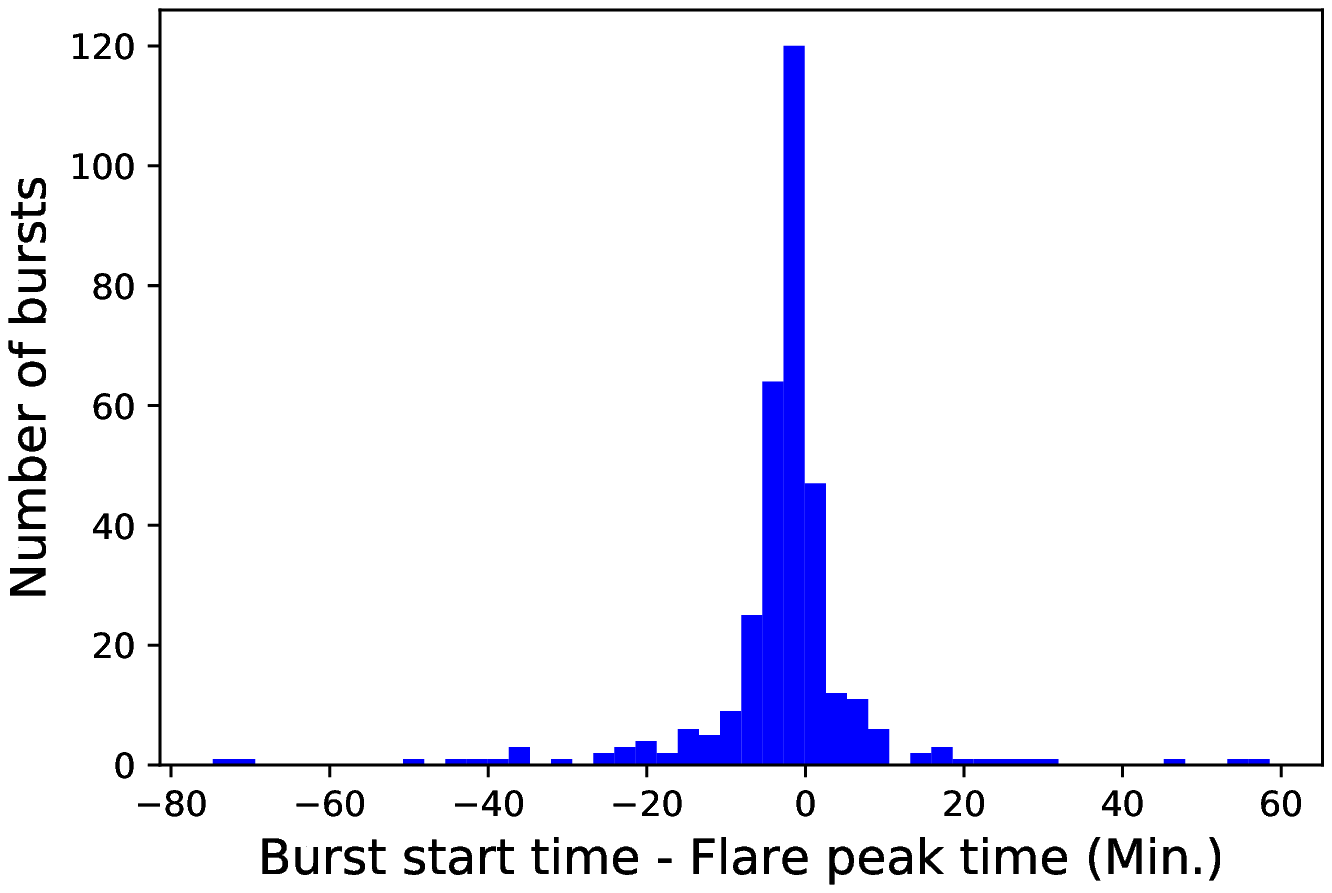}
\caption{In the left panel time difference between start time of burst and flare onset time vs number bursts is shown. In the right panel time difference between burst start time and the flare peak time vs number bursts is shown.}
\label{fig:timeplots}
\end{figure}

We have studied the relationship between onset, peak and end time of the flare and start time of the type III bursts. In Figure \ref{fig:timeplots}, left panel shows burst start time - flare onset time vs number of bursts. It is evident that most of the bursts in this sample are originated within 30 mins after onset of the flare. The right panel shows burst time - peak time of the flare vs number of bursts. It is evident from the plot that most of the type III bursts are occurred close to peak of the flare (approximately $\pm 10$ mins to the peak).

\begin{figure*}[!ht]
   \centerline{\hspace*{0.015\textwidth}
               \includegraphics[width=0.5\textwidth,clip=]{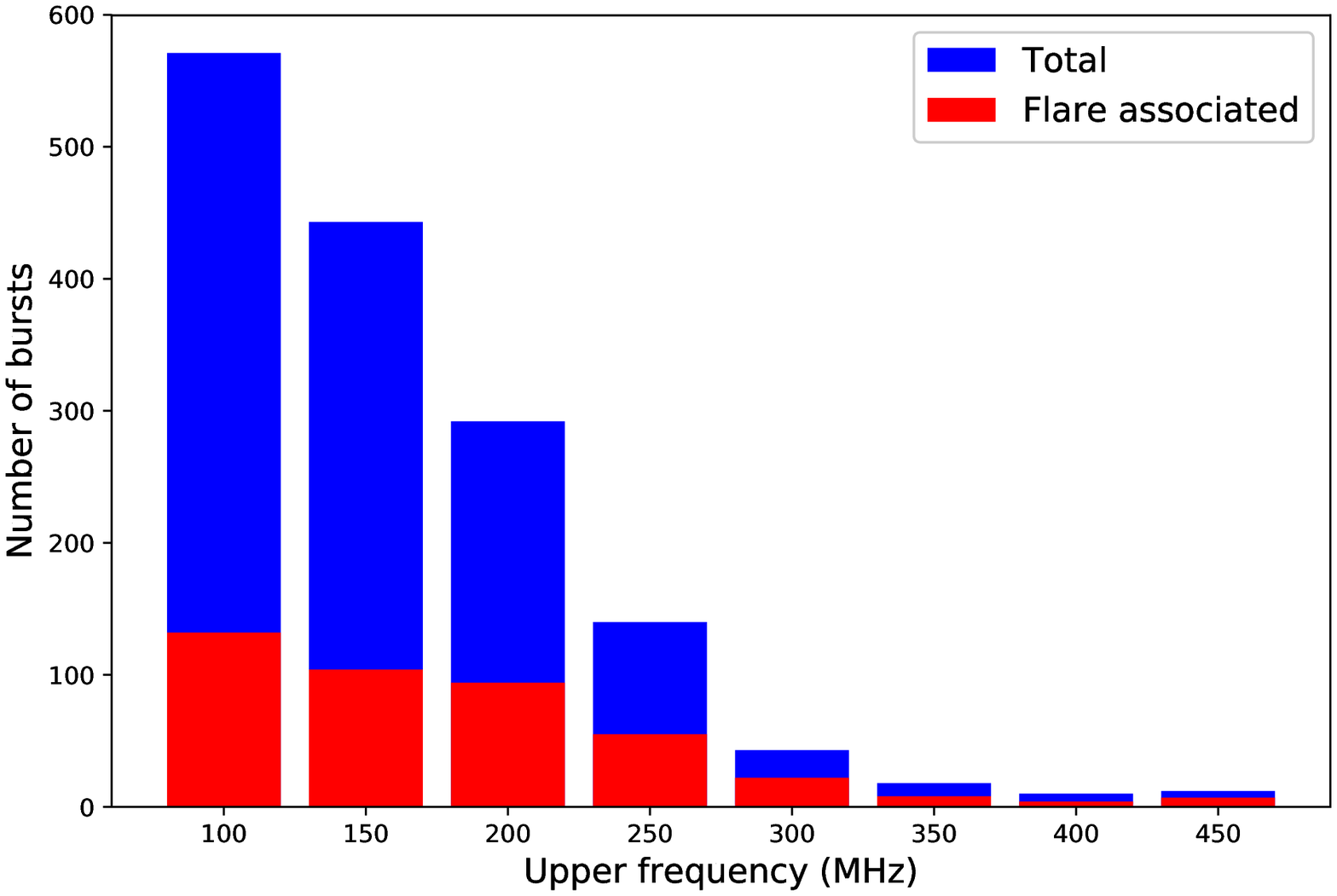}
               \hspace*{-0.06\textwidth}
               \includegraphics[width=0.5\textwidth,clip=]{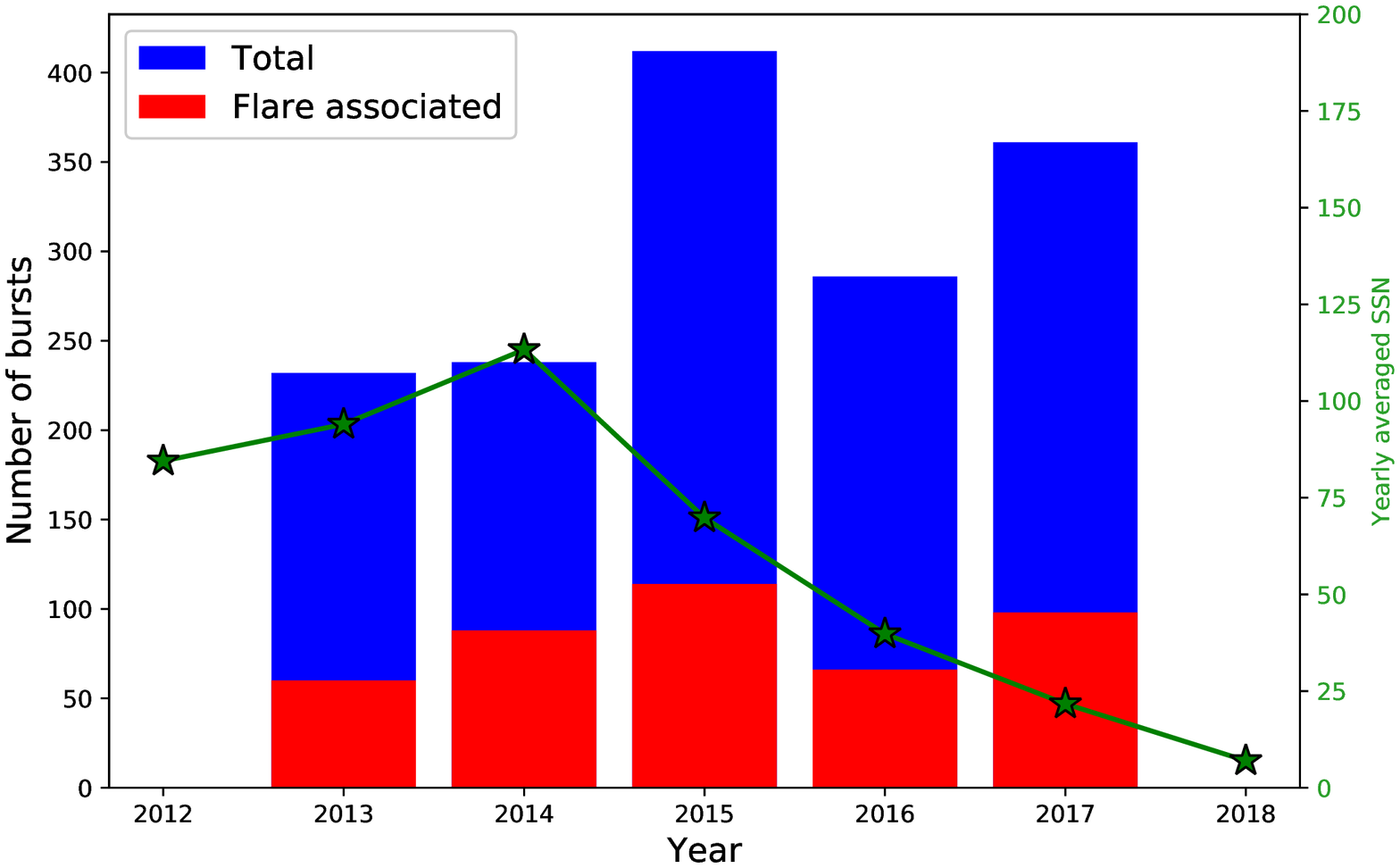}
              }
     \vspace{-0.24\textwidth}   
     \centerline{\Large \bf     
      \hspace{0.2 \textwidth}  \color{black}{(a)}
      \hspace{0.3\textwidth}  \color{black}{(b)}
         \hfill}
     \vspace{0.20\textwidth}    
   \centerline{\hspace*{0.015\textwidth}
               \includegraphics[width=0.5\textwidth,clip=]{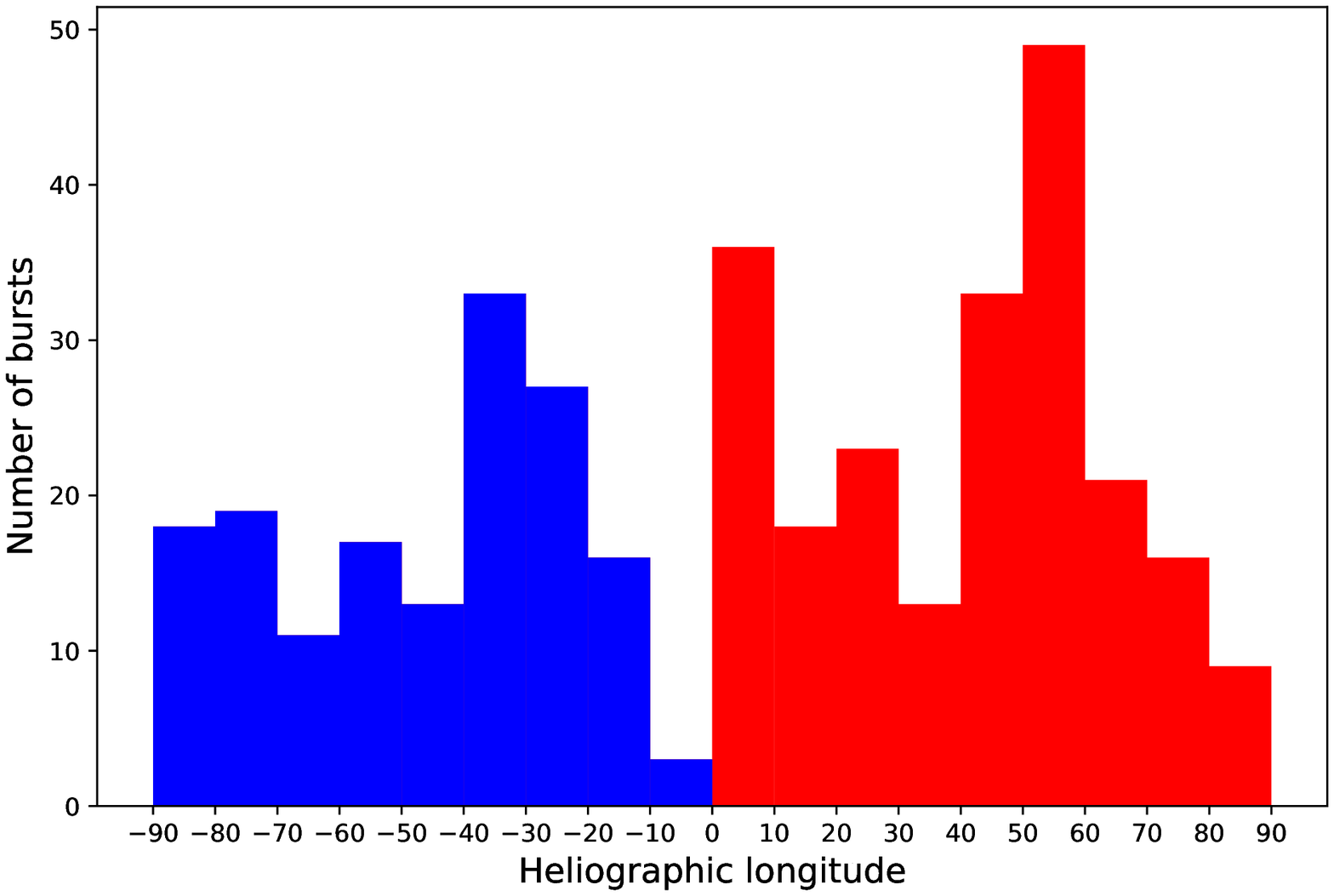}
               \hspace*{-0.06\textwidth}
               \includegraphics[width=0.5\textwidth,clip=]{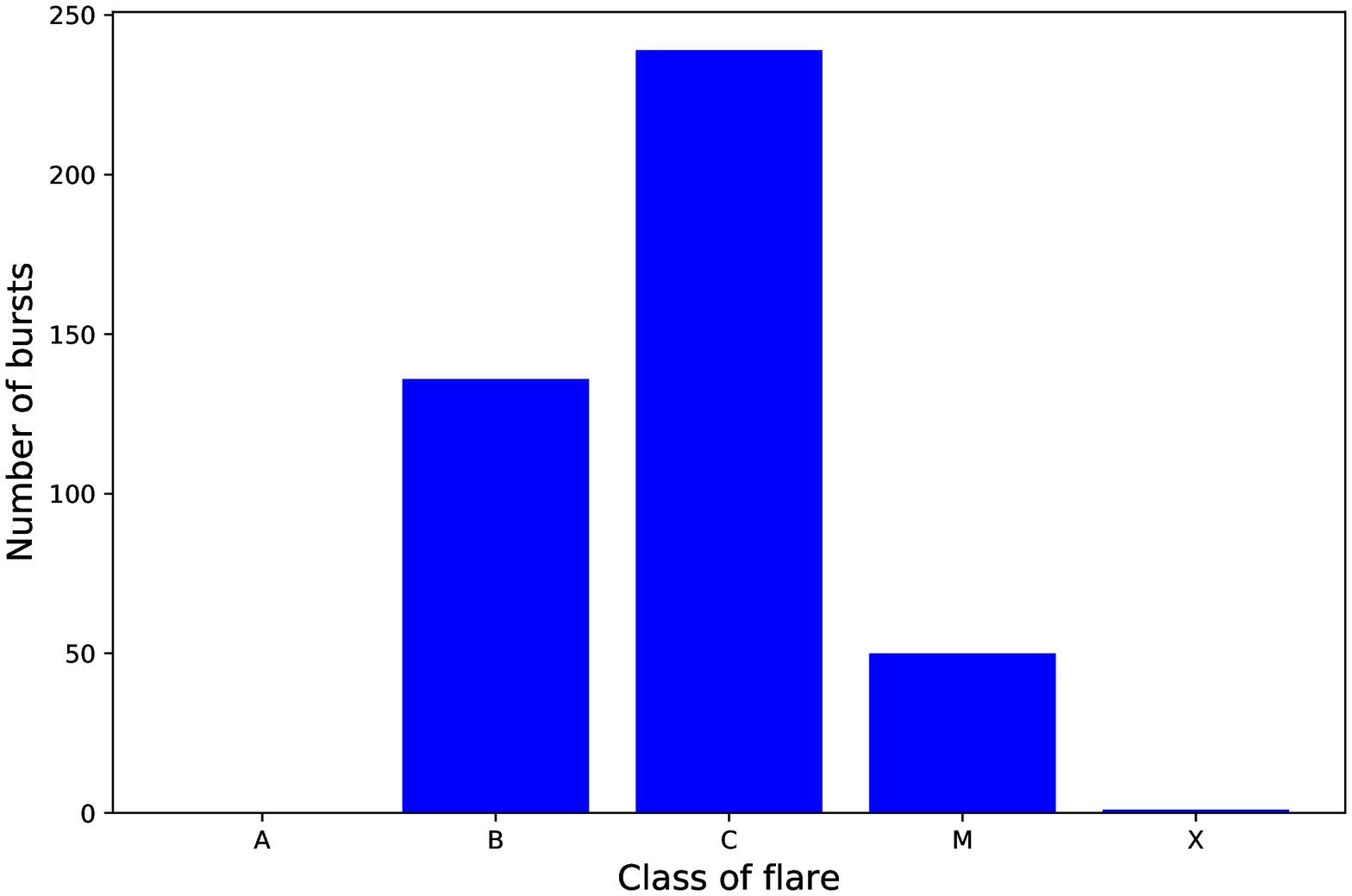}
              }
     \vspace{-0.26\textwidth}   
     \centerline{\Large \bf     
      \hspace{0.1 \textwidth} \color{black}{(c)}
      \hspace{0.39\textwidth}  \color{black}{(d)}
         \hfill}
     \vspace{0.22\textwidth}    
                
 \caption{The relationship between the parameters derived using radio dynamic spectrogram, light-curves from the GOES X-ray profiles and various instruments on board SDO. Panel (a) shows upper-frequency cutoff of radio bursts vs number of bursts. The blue and red colors indicate the total number of bursts and flare associated bursts respectively. Panel (b) shows the total number bursts and flare associated type III bursts observed in different years. The green markers indicate the yearly averaged sunspot number. Panel (c) shows total number of bursts observed in different heliographic longitudes. The blue and red colors indicates the east and west heliographic longitudes and panel (d) shows the class of flare vs the number of bursts.}
\label{fig:char}
\end{figure*}

In Figure \ref{fig:char}, panel (a) shows the upper frequency cut-off of radio bursts vs the number of observed radio bursts. The blue and red color bars in panel (a) indicate total number of bursts and flare associated bursts respectively. It is evident from the plot that both total number of bursts and flare associated bursts when plotted against upper frequency cutoff of bursts shows a power law. Panel (b) shows total number of bursts (in blue) and flare associated bursts (in red) that are observed in different years. We found that total number of bursts observed in different years weakly correlates with the yearly averaged sunspot number (the green curve). Note that we have used the sunspot number\footnote{\url{http://www.sidc.be/silso/datafiles}} provided by \citet{Cle2016}. Panel (c) shows variation of number of flare associated type III bursts with the heliographic longitudes. The blue and red histograms indicate the east and west longitudes respectively. This plot is evident that total number of flare associated bursts in the west longitudes are higher than the east. As previously mentioned this could be due to the fact that type III bursts possesses directivity and also Parker spirals that are originated in the west longitudes are directed towards the earth. In the east longitudes it is possible that although there exist equal number of bursts when compared to west longitudes, because of their directivity we could not observe part of bursts from the earth. Panel (d) shows the class of the flares classified based on GOES X ray flux vs the number of flare associated bursts. In this study, out of 426 flare association bursts, we have found 239 (56\%) bursts are due to C-class, 136 (32\%) bursts are due to B-class, 136 (32\%) bursts belongs to M-class and one bursts is due to X-class flare. We do not observe any A-class flare associated bursts, presumably because the sensitivity of CALLISTO spectrometer is not sufficient. From the plot, it can be noticed that most of the type III bursts are triggered by C-class flares. 

\section{Summary and Conclusions}
\label{Summ}

We have studied 1531 type III bursts observed in the frequency range 45 - 410 MHz during 2013 - 2017. The observations are carried out using CALLISTO spectrometer that is located at the Gauribidanur Radio Observatory, India. We have carried out the statistical study of the way upper and lower frequency cutoff vary with onset - peak - end times of flare associated type III bursts, flare class and location of the active region. Conclusions that are drawn from this study are given below. 

\begin{itemize}
\item We have found that only 28 $\%$ of bursts are flare associated and the remaining bursts may be originated from the weak energy release events (like jets) that are present in the solar corona. 

\item We found that most of the type III bursts in this sample are triggered by the C-class flares.

\item We found that most of the flare associated type III bursts are observed to be originated from the active regions whose heliographic latitudes are $\pm 23^\circ$. Note that this work is carried out during 2013 - 2017 (i.e. during solar cycle 24); it will be interesting to carry out similar studies at different phases of solar cycle and magnetic polarities.

\item We found that the bursts whose upper frequency is higher than 350 MHz are originated from the west longitudes. We interprete them that it could be due to the fact that Parker spirals from the west longitudes are directing towards the earth and type III bursts possesses the directivity. Also this work is corroborating the fact that high frequency type III bursts are more directive than low frequency bursts \citep{Singh2019}.

\item We found that most of the type III bursts are occurred within 30 minutes of the onset of the flare. 

\item It was observed that the most of the type III bursts are occurred during the peak time of the flare (with in $\pm 10$ mins from the peak of the flare).

\item We found that both total number of bursts and flare associated bursts show the power law when plotted against upper frequency cutoff of the type III bursts.

\item We found that yearly measured total number of type III burst and flare associated bursts weakly correlates with the yearly averaged sunspot numbers.

\item This study infers that number of bursts observed from the west longitudes are larger than the ones from the east longitudes. Presumably, some of the bursts originated in the east longitudes doesn't reach earth because of their high directivity. 

\end{itemize}

\acknowledgments
The sunspot number used in this article is credited to WDC-SILSO, Royal Observatory of Belgium, Brussels. This research used version 2.0.1 of the SunPy open source software package \citep{sunpy_community2020}. We thank the data centre of the e-Callisto network which is hosted by the FHNW, Institute for Data Science, Switzerland. We thank the referee for his constructive suggestions of the manuscript.
\vspace{5mm}

\bibliography{ms}{}

\begin{thebibliography}{33}
\ifx\bisbn     \undefined \def\bisbn  #1{ISBN #1}\fi
\ifx\binits    \undefined \def\binits#1{#1}\fi
\ifx\bauthor   \undefined \def\bauthor#1{#1}\fi
\ifx\batitle   \undefined \def\batitle#1{#1}\fi
\ifx\bjtitle   \undefined \def\bjtitle#1{\textit{#1}}\fi
\ifx\bvolume   \undefined \def\bvolume#1{\textbf{#1}}\fi
\ifx\byear     \undefined \def\byear#1{#1}\fi
\ifx\bissue    \undefined \def\bissue#1{#1}\fi
\ifx\bfpage    \undefined \def\bfpage#1{#1}\fi
\ifx\blpage    \undefined \def\blpage #1{#1}\fi
\ifx\burl      \undefined \def\burl#1{\textsf{#1}}\fi
\ifx\href      \undefined \def\href#1#2{\textsf{#2}}\fi
\ifx\betal     \undefined \def\betal{\textit{et al.}}\fi
\ifx\bctitle   \undefined \def\bctitle#1{#1}\fi
\ifx\beditor   \undefined \def\beditor#1{#1}\fi
\ifx\bbtitle   \undefined \def\bbtitle#1{\textit{#1}}\fi
\ifx\bedition  \undefined \def\bedition#1{#1}\fi
\ifx\bseriesno \undefined \def\bseriesno#1{\textbf{#1}}\fi
\ifx\blocation \undefined \def\blocation#1{#1}\fi
\ifx\bsertitle \undefined \def\bsertitle#1{\textit{#1}}\fi
\ifx\bsnm      \undefined \def\bsnm#1{#1}\fi
\ifx\bsuffix   \undefined \def\bsuffix#1{#1}\fi
\ifx\bparticle \undefined \def\bparticle#1{#1}\fi
\ifx\barticle  \undefined \def\barticle#1{}\fi
\ifx\binstitute  \undefined \def\binstitute#1{#1}\fi
\ifx\bpublisher  \undefined \def\bpublisher#1{#1}\fi
\ifx\doiurl    \undefined
  \def\doiurl#1{\href{http://dx.doi.org/#1}{\textsf{DOI}}}\fi
\ifx\arxivurl  \undefined
  \def\arxivurl#1{\href{http://arxiv.org/abs/#1}{\textsf{arXiv}}}\fi
\ifx\adsurl    \undefined
  \def\adsurl#1{\href{http://adsabs.harvard.edu/abs/#1}{\textsf{ADS}}}\fi
\ifx\botherref \undefined \def\botherref#1{}\fi
\ifx\url       \undefined \def\url#1{\textsf{#1}}\fi
\ifx\bchapter  \undefined \def\bchapter#1{}\fi
\ifx\bbook     \undefined \def\bbook#1{}\fi
\ifx\bcomment  \undefined \def\bcomment#1{#1}\fi
\ifx\oauthor   \undefined \def\oauthor#1{#1}\fi
\ifx\citeauthoryear \undefined\def \citeauthoryear#1{#1}\fi
\ifx\endbibitem\undefined \def\endbibitem{}\fi
\ifx\bconflocation  \undefined \def\bconflocation#1{#1} \fi

\bibitem[\protect\citeauthoryear{{Benz} \textit{et~al.}}{2009}]{Benz2009}
\begin{barticle}
\bauthor{\bsnm{{Benz}}, \binits{A.O.}},
\bauthor{\bsnm{{Monstein}}, \binits{C.}},
\bauthor{\bsnm{{Meyer}}, \binits{H.}},
\bauthor{\bsnm{{Manoharan}}, \binits{P.K.}},
\bauthor{\bsnm{{Ramesh}}, \binits{R.}},
\bauthor{\bsnm{{Altyntsev}}, \binits{A.}},
\bauthor{\bsnm{{Lara}}, \binits{A.}},
\bauthor{\bsnm{{Paez}}, \binits{J.}},
\bauthor{\bsnm{{Cho}}, \binits{K.-S.}}:
\byear{2009},
\batitle{{A World-Wide Net of Solar Radio Spectrometers: e-CALLISTO}}.
\bjtitle{Earth Moon and Planets}
\bvolume{104}(\bissue{1-4}),
\bfpage{277}.
\doiurl{10.1007/s11038-008-9267-6}.
\adsurl{https://ui.adsabs.harvard.edu/abs/2009EM&P..104..277B}.
\end{barticle}
\endbibitem

\bibitem[\protect\citeauthoryear{{Clette} \textit{et~al.}}{2016}]{Cle2016}
\begin{barticle}
\bauthor{\bsnm{{Clette}}, \binits{F.}},
\bauthor{\bsnm{{Lef{\`e}vre}}, \binits{L.}},
\bauthor{\bsnm{{Cagnotti}}, \binits{M.}},
\bauthor{\bsnm{{Cortesi}}, \binits{S.}},
\bauthor{\bsnm{{Bulling}}, \binits{A.}}:
\byear{2016},
\batitle{{The Revised Brussels-Locarno Sunspot Number (1981 - 2015)}}.
\bjtitle{\solphys}
\bvolume{291}(\bissue{9-10}),
\bfpage{2733}.
\doiurl{10.1007/s11207-016-0875-4}.
\adsurl{https://ui.adsabs.harvard.edu/abs/2016SoPh..291.2733C}.
\end{barticle}
\endbibitem

\bibitem[\protect\citeauthoryear{{Ginzburg} and
  {Zhelezniakov}}{1958}]{Ginzburg1958}
\begin{barticle}
\bauthor{\bsnm{{Ginzburg}}, \binits{V.L.}},
\bauthor{\bsnm{{Zhelezniakov}}, \binits{V.V.}}:
\byear{1958},
\batitle{{On the Possible Mechanisms of Sporadic Solar Radio Emission
  (Radiation in an Isotropic Plasma)}}.
\bjtitle{\sovast}
\bvolume{2},
\bfpage{653}.
\adsurl{https://ui.adsabs.harvard.edu/abs/1958SvA.....2..653G}.
\end{barticle}
\endbibitem

\bibitem[\protect\citeauthoryear{{James} and {Subramanian}}{2018}]{James2018}
\begin{barticle}
\bauthor{\bsnm{{James}}, \binits{T.}},
\bauthor{\bsnm{{Subramanian}}, \binits{P.}}:
\byear{2018},
\batitle{{Energetics of small electron acceleration episodes in the solar
  corona from radio noise storm observations}}.
\bjtitle{\mnras}
\bvolume{479}(\bissue{2}),
\bfpage{1603}.
\doiurl{10.1093/mnras/sty1216}.
\adsurl{https://ui.adsabs.harvard.edu/abs/2018MNRAS.479.1603J}.
\end{barticle}
\endbibitem

\bibitem[\protect\citeauthoryear{{James}, {Subramanian}, and
  {Kontar}}{2017}]{James2017}
\begin{barticle}
\bauthor{\bsnm{{James}}, \binits{T.}},
\bauthor{\bsnm{{Subramanian}}, \binits{P.}},
\bauthor{\bsnm{{Kontar}}, \binits{E.P.}}:
\byear{2017},
\batitle{{Small electron acceleration episodes in the solar corona}}.
\bjtitle{\mnras}
\bvolume{471}(\bissue{1}),
\bfpage{89}.
\doiurl{10.1093/mnras/stx1460}.
\adsurl{https://ui.adsabs.harvard.edu/abs/2017MNRAS.471...89J}.
\end{barticle}
\endbibitem

\bibitem[\protect\citeauthoryear{{Kerdraon} and {Delouis}}{1997}]{Ker1997}
\begin{bbook}
\bauthor{\bsnm{{Kerdraon}}, \binits{A.}},
\bauthor{\bsnm{{Delouis}}, \binits{J.-M.}}:
\byear{1997},
In: \beditor{\bsnm{{Trottet}}, \binits{G.}} (ed.)
\bbtitle{{The Nan{\c{c}}ay Radioheliograph}}
\bseriesno{483},
\bfpage{192}.
\doiurl{10.1007/BFb0106458}.
\adsurl{https://ui.adsabs.harvard.edu/abs/1997LNP...483..192K}.
\end{bbook}
\endbibitem

\bibitem[\protect\citeauthoryear{{Kishore} \textit{et~al.}}{2015}]{Kishore2015}
\begin{barticle}
\bauthor{\bsnm{{Kishore}}, \binits{P.}},
\bauthor{\bsnm{{Ramesh}}, \binits{R.}},
\bauthor{\bsnm{{Kathiravan}}, \binits{C.}},
\bauthor{\bsnm{{Rajalingam}}, \binits{M.}}:
\byear{2015},
\batitle{{A Low-Frequency Radio Spectropolarimeter for Observations of the
  Solar Corona}}.
\bjtitle{\solphys}
\bvolume{290}(\bissue{9}),
\bfpage{2409}.
\doiurl{10.1007/s11207-015-0705-0}.
\adsurl{https://ui.adsabs.harvard.edu/abs/2015SoPh..290.2409K}.
\end{barticle}
\endbibitem

\bibitem[\protect\citeauthoryear{{Kishore} \textit{et~al.}}{2017}]{Kishore2017}
\begin{barticle}
\bauthor{\bsnm{{Kishore}}, \binits{P.}},
\bauthor{\bsnm{{Kathiravan}}, \binits{C.}},
\bauthor{\bsnm{{Ramesh}}, \binits{R.}},
\bauthor{\bsnm{{Ebenezer}}, \binits{E.}}:
\byear{2017},
\batitle{{Coronal Magnetic Field Lines and Electrons Associated with Type III-V
  Radio Bursts in a Solar Flare}}.
\bjtitle{Journal of Astrophysics and Astronomy}
\bvolume{38}(\bissue{2}),
\bfpage{24}.
\doiurl{10.1007/s12036-017-9444-y}.
\adsurl{https://ui.adsabs.harvard.edu/abs/2017JApA...38...24K}.
\end{barticle}
\endbibitem

\bibitem[\protect\citeauthoryear{{Lemen} \textit{et~al.}}{2012}]{Lem2012}
\begin{barticle}
\bauthor{\bsnm{{Lemen}}, \binits{J.R.}},
\bauthor{\bsnm{{Title}}, \binits{A.M.}},
\bauthor{\bsnm{{Akin}}, \binits{D.J.}},
\bauthor{\bsnm{{Boerner}}, \binits{P.F.}},
\bauthor{\bsnm{{Chou}}, \binits{C.}},
\bauthor{\bsnm{{Drake}}, \binits{J.F.}},
\bauthor{\bsnm{{Duncan}}, \binits{D.W.}},
\bauthor{\bsnm{{Edwards}}, \binits{C.G.}},
\bauthor{\bsnm{{Friedlaender}}, \binits{F.M.}},
\bauthor{\bsnm{{Heyman}}, \binits{G.F.}},
\bauthor{\bsnm{{Hurlburt}}, \binits{N.E.}},
\bauthor{\bsnm{{Katz}}, \binits{N.L.}},
\bauthor{\bsnm{{Kushner}}, \binits{G.D.}},
\bauthor{\bsnm{{Levay}}, \binits{M.}},
\bauthor{\bsnm{{Lindgren}}, \binits{R.W.}},
\bauthor{\bsnm{{Mathur}}, \binits{D.P.}},
\bauthor{\bsnm{{McFeaters}}, \binits{E.L.}},
\bauthor{\bsnm{{Mitchell}}, \binits{S.}},
\bauthor{\bsnm{{Rehse}}, \binits{R.A.}},
\bauthor{\bsnm{{Schrijver}}, \binits{C.J.}},
\bauthor{\bsnm{{Springer}}, \binits{L.A.}},
\bauthor{\bsnm{{Stern}}, \binits{R.A.}},
\bauthor{\bsnm{{Tarbell}}, \binits{T.D.}},
\bauthor{\bsnm{{Wuelser}}, \binits{J.-P.}},
\bauthor{\bsnm{{Wolfson}}, \binits{C.J.}},
\bauthor{\bsnm{{Yanari}}, \binits{C.}},
\bauthor{\bsnm{{Bookbinder}}, \binits{J.A.}},
\bauthor{\bsnm{{Cheimets}}, \binits{P.N.}},
\bauthor{\bsnm{{Caldwell}}, \binits{D.}},
\bauthor{\bsnm{{Deluca}}, \binits{E.E.}},
\bauthor{\bsnm{{Gates}}, \binits{R.}},
\bauthor{\bsnm{{Golub}}, \binits{L.}},
\bauthor{\bsnm{{Park}}, \binits{S.}},
\bauthor{\bsnm{{Podgorski}}, \binits{W.A.}},
\bauthor{\bsnm{{Bush}}, \binits{R.I.}},
\bauthor{\bsnm{{Scherrer}}, \binits{P.H.}},
\bauthor{\bsnm{{Gummin}}, \binits{M.A.}},
\bauthor{\bsnm{{Smith}}, \binits{P.}},
\bauthor{\bsnm{{Auker}}, \binits{G.}},
\bauthor{\bsnm{{Jerram}}, \binits{P.}},
\bauthor{\bsnm{{Pool}}, \binits{P.}},
\bauthor{\bsnm{{Soufli}}, \binits{R.}},
\bauthor{\bsnm{{Windt}}, \binits{D.L.}},
\bauthor{\bsnm{{Beardsley}}, \binits{S.}},
\bauthor{\bsnm{{Clapp}}, \binits{M.}},
\bauthor{\bsnm{{Lang}}, \binits{J.}},
\bauthor{\bsnm{{Waltham}}, \binits{N.}}:
\byear{2012},
\batitle{{The Atmospheric Imaging Assembly (AIA) on the Solar Dynamics
  Observatory (SDO)}}.
\bjtitle{\solphys}
\bvolume{275}(\bissue{1-2}),
\bfpage{17}.
\doiurl{10.1007/s11207-011-9776-8}.
\adsurl{https://ui.adsabs.harvard.edu/abs/2012SoPh..275...17L}.
\end{barticle}
\endbibitem

\bibitem[\protect\citeauthoryear{{Melrose}}{1980}]{Mel1980}
\begin{barticle}
\bauthor{\bsnm{{Melrose}}, \binits{D.B.}}:
\byear{1980},
\batitle{{The Emission Mechanisms for Solar Radio Bursts}}.
\bjtitle{\ssr}
\bvolume{26}(\bissue{1}),
\bfpage{3}.
\doiurl{10.1007/BF00212597}.
\adsurl{https://ui.adsabs.harvard.edu/abs/1980SSRv...26....3M}.
\end{barticle}
\endbibitem

\bibitem[\protect\citeauthoryear{{Monstein}, {Ramesh}, and
  {Kathiravan}}{2007}]{Mon2007}
\begin{barticle}
\bauthor{\bsnm{{Monstein}}, \binits{C.}},
\bauthor{\bsnm{{Ramesh}}, \binits{R.}},
\bauthor{\bsnm{{Kathiravan}}, \binits{C.}}:
\byear{2007},
\batitle{{Radio spectrum measurements at the Gauribidanur observatory}}.
\bjtitle{Bulletin of the Astronomical Society of India}
\bvolume{35},
\bfpage{473}.
\adsurl{https://ui.adsabs.harvard.edu/abs/2007BASI...35..473M}.
\end{barticle}
\endbibitem

\bibitem[\protect\citeauthoryear{{Mugundhan}
  \textit{et~al.}}{2016}]{Mugundhan2016}
\begin{barticle}
\bauthor{\bsnm{{Mugundhan}}, \binits{V.}},
\bauthor{\bsnm{{Ramesh}}, \binits{R.}},
\bauthor{\bsnm{{Barve}}, \binits{I.V.}},
\bauthor{\bsnm{{Kathiravan}}, \binits{C.}},
\bauthor{\bsnm{{Gireesh}}, \binits{G.V.S.}},
\bauthor{\bsnm{{Kharb}}, \binits{P.}},
\bauthor{\bsnm{{Misra}}, \binits{A.}}:
\byear{2016},
\batitle{{Low-Frequency Radio Observations of the Solar Corona with Arcminute
  Angular Resolution: Implications for Coronal Turbulence and Weak Energy
  Releases}}.
\bjtitle{\apj}
\bvolume{831}(\bissue{2}),
\bfpage{154}.
\doiurl{10.3847/0004-637X/831/2/154}.
\adsurl{https://ui.adsabs.harvard.edu/abs/2016ApJ...831..154M}.
\end{barticle}
\endbibitem

\bibitem[\protect\citeauthoryear{{Mugundhan}
  \textit{et~al.}}{2018}]{Mugundhan2018}
\begin{barticle}
\bauthor{\bsnm{{Mugundhan}}, \binits{V.}},
\bauthor{\bsnm{{Ramesh}}, \binits{R.}},
\bauthor{\bsnm{{Kathiravan}}, \binits{C.}},
\bauthor{\bsnm{{Gireesh}}, \binits{G.V.S.}},
\bauthor{\bsnm{{Kumari}}, \binits{A.}},
\bauthor{\bsnm{{Hariharan}}, \binits{K.}},
\bauthor{\bsnm{{Barve}}, \binits{I.V.}}:
\byear{2018},
\batitle{{The First Low-frequency Radio Observations of the Solar Corona on
  {\ensuremath{\approx}}200 km Long Interferometer Baseline}}.
\bjtitle{\apjl}
\bvolume{855}(\bissue{1}),
\bfpage{L8}.
\doiurl{10.3847/2041-8213/aaaf64}.
\adsurl{https://ui.adsabs.harvard.edu/abs/2018ApJ...855L...8M}.
\end{barticle}
\endbibitem

\bibitem[\protect\citeauthoryear{{Pesnell}, {Thompson}, and
  {Chamberlin}}{2012}]{Pes2012}
\begin{barticle}
\bauthor{\bsnm{{Pesnell}}, \binits{W.D.}},
\bauthor{\bsnm{{Thompson}}, \binits{B.J.}},
\bauthor{\bsnm{{Chamberlin}}, \binits{P.C.}}:
\byear{2012},
\batitle{{The Solar Dynamics Observatory (SDO)}}.
\bjtitle{\solphys}
\bvolume{275}(\bissue{1-2}),
\bfpage{3}.
\doiurl{10.1007/s11207-011-9841-3}.
\adsurl{https://ui.adsabs.harvard.edu/abs/2012SoPh..275....3P}.
\end{barticle}
\endbibitem

\bibitem[\protect\citeauthoryear{{Ramesh}}{2011}]{Ramesh2011}
\begin{bchapter}
\bauthor{\bsnm{{Ramesh}}, \binits{R.}}:
\byear{2011},
\bctitle{{Low frequency solar radio astronomy at the Indian Institute of
  Astrophysics (IIA)}}.
In: \bbtitle{Astronomical Society of India Conference Series},
\bsertitle{Astronomical Society of India Conference Series}
\bseriesno{2},
\bfpage{55}.
\adsurl{https://ui.adsabs.harvard.edu/abs/2011ASInC...2...55R}.
\end{bchapter}
\endbibitem

\bibitem[\protect\citeauthoryear{{Ramesh}}{2014}]{Ramesh2014}
\begin{bchapter}
\bauthor{\bsnm{{Ramesh}}, \binits{R.}}:
\byear{2014},
\bctitle{{Solar observations at low frequencies with the Gauribidanur
  radioheliograph}}.
In: \beditor{\bsnm{{Chengalur}}, \binits{J.N.}},
\beditor{\bsnm{{Gupta}}, \binits{Y.}} (eds.)
\bbtitle{Metrewavelength Sky},
\bsertitle{Astron. Soc. India Conf. Ser.}
\bseriesno{13},
\bfpage{19}.
\adsurl{https://ui.adsabs.harvard.edu/abs/2014ASInC..13...19R/abstract}.
\end{bchapter}
\endbibitem

\bibitem[\protect\citeauthoryear{{Ramesh} \textit{et~al.}}{2003}]{Ramesh2003}
\begin{barticle}
\bauthor{\bsnm{{Ramesh}}, \binits{R.}},
\bauthor{\bsnm{{Kathiravan}}, \binits{C.}},
\bauthor{\bsnm{{Narayanan}}, \binits{A.S.}},
\bauthor{\bsnm{{Ebenezer}}, \binits{E.}}:
\byear{2003},
\batitle{{Metric observations of transient, quasi-periodic radio emission from
  the solar corona in association with a ``halo`` CME and an ``EIT wave''
  event}}.
\bjtitle{\aap}
\bvolume{400},
\bfpage{753}.
\doiurl{10.1051/0004-6361:20030019}.
\adsurl{https://ui.adsabs.harvard.edu/abs/2003A&A...400..753R}.
\end{barticle}
\endbibitem

\bibitem[\protect\citeauthoryear{{Ramesh} \textit{et~al.}}{2005}]{Ramesh2005}
\begin{barticle}
\bauthor{\bsnm{{Ramesh}}, \binits{R.}},
\bauthor{\bsnm{{Narayanan}}, \binits{A.S.}},
\bauthor{\bsnm{{Kathiravan}}, \binits{C.}},
\bauthor{\bsnm{{Sastry}}, \binits{C.V.}},
\bauthor{\bsnm{{Shankar}}, \binits{N.U.}}:
\byear{2005},
\batitle{{An estimation of the plasma parameters in the solar corona using
  quasi-periodic metric type III radio burst emission}}.
\bjtitle{\aap}
\bvolume{431},
\bfpage{353}.
\doiurl{10.1051/0004-6361:20041130}.
\adsurl{https://ui.adsabs.harvard.edu/abs/2005A&A...431..353R}.
\end{barticle}
\endbibitem

\bibitem[\protect\citeauthoryear{{Ramesh} \textit{et~al.}}{2010}]{Ram2010}
\begin{barticle}
\bauthor{\bsnm{{Ramesh}}, \binits{R.}},
\bauthor{\bsnm{{Kathiravan}}, \binits{C.}},
\bauthor{\bsnm{{Barve}}, \binits{I.V.}},
\bauthor{\bsnm{{Beeharry}}, \binits{G.K.}},
\bauthor{\bsnm{{Rajasekara}}, \binits{G.N.}}:
\byear{2010},
\batitle{{Radio Observations of Weak Energy Releases in the Solar Corona}}.
\bjtitle{\apjl}
\bvolume{719}(\bissue{1}),
\bfpage{L41}.
\doiurl{10.1088/2041-8205/719/1/L41}.
\adsurl{https://ui.adsabs.harvard.edu/abs/2010ApJ...719L..41R}.
\end{barticle}
\endbibitem

\bibitem[\protect\citeauthoryear{{Ramesh} \textit{et~al.}}{2013}]{Ram2013}
\begin{barticle}
\bauthor{\bsnm{{Ramesh}}, \binits{R.}},
\bauthor{\bsnm{{Sasikumar Raja}}, \binits{K.}},
\bauthor{\bsnm{{Kathiravan}}, \binits{C.}},
\bauthor{\bsnm{{Narayanan}}, \binits{A.S.}}:
\byear{2013},
\batitle{{Low-frequency Radio Observations of Picoflare Category Energy
  Releases in the Solar Atmosphere}}.
\bjtitle{\apj}
\bvolume{762}(\bissue{2}),
\bfpage{89}.
\doiurl{10.1088/0004-637X/762/2/89}.
\adsurl{https://ui.adsabs.harvard.edu/abs/2013ApJ...762...89R}.
\end{barticle}
\endbibitem

\bibitem[\protect\citeauthoryear{{Reid} and {Ratcliffe}}{2014}]{Rei2014}
\begin{barticle}
\bauthor{\bsnm{{Reid}}, \binits{H.A.S.}},
\bauthor{\bsnm{{Ratcliffe}}, \binits{H.}}:
\byear{2014},
\batitle{{A review of solar type III radio bursts}}.
\bjtitle{Research in Astronomy and Astrophysics}
\bvolume{14}(\bissue{7}),
\bfpage{773}.
\doiurl{10.1088/1674-4527/14/7/003}.
\adsurl{https://ui.adsabs.harvard.edu/abs/2014RAA....14..773R}.
\end{barticle}
\endbibitem

\bibitem[\protect\citeauthoryear{{Saint-Hilaire}, {Vilmer}, and
  {Kerdraon}}{2013}]{Saint2013}
\begin{barticle}
\bauthor{\bsnm{{Saint-Hilaire}}, \binits{P.}},
\bauthor{\bsnm{{Vilmer}}, \binits{N.}},
\bauthor{\bsnm{{Kerdraon}}, \binits{A.}}:
\byear{2013},
\batitle{{A Decade of Solar Type III Radio Bursts Observed by the Nan{\c{c}}ay
  Radioheliograph 1998-2008}}.
\bjtitle{\apj}
\bvolume{762}(\bissue{1}),
\bfpage{60}.
\doiurl{10.1088/0004-637X/762/1/60}.
\adsurl{https://ui.adsabs.harvard.edu/abs/2013ApJ...762...60S}.
\end{barticle}
\endbibitem

\bibitem[\protect\citeauthoryear{{Sasikumar Raja} and {Ramesh}}{2013}]{Sas2013}
\begin{barticle}
\bauthor{\bsnm{{Sasikumar Raja}}, \binits{K.}},
\bauthor{\bsnm{{Ramesh}}, \binits{R.}}:
\byear{2013},
\batitle{{Low-frequency Observations of Transient Quasi-periodic Radio Emission
  from the Solar Atmosphere}}.
\bjtitle{\apj}
\bvolume{775}(\bissue{1}),
\bfpage{38}.
\doiurl{10.1088/0004-637X/775/1/38}.
\adsurl{https://ui.adsabs.harvard.edu/abs/2013ApJ...775...38S}.
\end{barticle}
\endbibitem

\bibitem[\protect\citeauthoryear{{Sasikumar Raja}
  \textit{et~al.}}{2018}]{Sas2018}
\begin{botherref}
\oauthor{\bsnm{{Sasikumar Raja}}, \binits{K.}},
\oauthor{\bsnm{{Subramanian}}, \binits{P.}},
\oauthor{\bsnm{{Ananthakrishnan}}, \binits{S.}},
\oauthor{\bsnm{{Monstein}}, \binits{C.}}:
2018,
{CALLISTO Spectrometer at IISER-Pune}.
\textit{arXiv e-prints},
arXiv:1801.03547.
\adsurl{https://ui.adsabs.harvard.edu/abs/2018arXiv180103547S}.
\end{botherref}
\endbibitem

\bibitem[\protect\citeauthoryear{{Schou} \textit{et~al.}}{2012}]{Sch2012}
\begin{barticle}
\bauthor{\bsnm{{Schou}}, \binits{J.}},
\bauthor{\bsnm{{Scherrer}}, \binits{P.H.}},
\bauthor{\bsnm{{Bush}}, \binits{R.I.}},
\bauthor{\bsnm{{Wachter}}, \binits{R.}},
\bauthor{\bsnm{{Couvidat}}, \binits{S.}},
\bauthor{\bsnm{{Rabello-Soares}}, \binits{M.C.}},
\bauthor{\bsnm{{Bogart}}, \binits{R.S.}},
\bauthor{\bsnm{{Hoeksema}}, \binits{J.T.}},
\bauthor{\bsnm{{Liu}}, \binits{Y.}},
\bauthor{\bsnm{{Duvall}}, \binits{T.L.}},
\bauthor{\bsnm{{Akin}}, \binits{D.J.}},
\bauthor{\bsnm{{Allard}}, \binits{B.A.}},
\bauthor{\bsnm{{Miles}}, \binits{J.W.}},
\bauthor{\bsnm{{Rairden}}, \binits{R.}},
\bauthor{\bsnm{{Shine}}, \binits{R.A.}},
\bauthor{\bsnm{{Tarbell}}, \binits{T.D.}},
\bauthor{\bsnm{{Title}}, \binits{A.M.}},
\bauthor{\bsnm{{Wolfson}}, \binits{C.J.}},
\bauthor{\bsnm{{Elmore}}, \binits{D.F.}},
\bauthor{\bsnm{{Norton}}, \binits{A.A.}},
\bauthor{\bsnm{{Tomczyk}}, \binits{S.}}:
\byear{2012},
\batitle{{Design and Ground Calibration of the Helioseismic and Magnetic Imager
  (HMI) Instrument on the Solar Dynamics Observatory (SDO)}}.
\bjtitle{\solphys}
\bvolume{275}(\bissue{1-2}),
\bfpage{229}.
\doiurl{10.1007/s11207-011-9842-2}.
\adsurl{https://ui.adsabs.harvard.edu/abs/2012SoPh..275..229S}.
\end{barticle}
\endbibitem

\bibitem[\protect\citeauthoryear{{Sharma}, {Oberoi}, and
  {Arjunwadkar}}{2018}]{Rohit2018}
\begin{barticle}
\bauthor{\bsnm{{Sharma}}, \binits{R.}},
\bauthor{\bsnm{{Oberoi}}, \binits{D.}},
\bauthor{\bsnm{{Arjunwadkar}}, \binits{M.}}:
\byear{2018},
\batitle{{Quantifying Weak Nonthermal Solar Radio Emission at Low Radio
  Frequencies}}.
\bjtitle{\apj}
\bvolume{852}(\bissue{2}),
\bfpage{69}.
\doiurl{10.3847/1538-4357/aa9d96}.
\adsurl{https://ui.adsabs.harvard.edu/abs/2018ApJ...852...69S}.
\end{barticle}
\endbibitem

\bibitem[\protect\citeauthoryear{{Singh} \textit{et~al.}}{2019}]{Singh2019}
\begin{barticle}
\bauthor{\bsnm{{Singh}}, \binits{D.}},
\bauthor{\bsnm{{Sasikumar Raja}}, \binits{K.}},
\bauthor{\bsnm{{Subramanian}}, \binits{P.}},
\bauthor{\bsnm{{Ramesh}}, \binits{R.}},
\bauthor{\bsnm{{Monstein}}, \binits{C.}}:
\byear{2019},
\batitle{{Automated Detection of Solar Radio Bursts Using a Statistical
  Method}}.
\bjtitle{\solphys}
\bvolume{294}(\bissue{8}),
\bfpage{112}.
\doiurl{10.1007/s11207-019-1500-0}.
\adsurl{https://ui.adsabs.harvard.edu/abs/2019SoPh..294..112S}.
\end{barticle}
\endbibitem

\bibitem[\protect\citeauthoryear{{Stewart}}{1974}]{Ste1974}
\begin{barticle}
\bauthor{\bsnm{{Stewart}}, \binits{R.T.}}:
\byear{1974},
\batitle{{Harmonic Ratios of Inverted-U Type III Bursts}}.
\bjtitle{\solphys}
\bvolume{39}(\bissue{2}),
\bfpage{451}.
\doiurl{10.1007/BF00162437}.
\adsurl{https://ui.adsabs.harvard.edu/abs/1974SoPh...39..451S}.
\end{barticle}
\endbibitem

\bibitem[\protect\citeauthoryear{{Suzuki} and {Sheridan}}{1982}]{Suz1982}
\begin{barticle}
\bauthor{\bsnm{{Suzuki}}, \binits{S.}},
\bauthor{\bsnm{{Sheridan}}, \binits{K.V.}}:
\byear{1982},
\batitle{{On the fundamental and harmonic components of low-frequency Type III
  solar radio bursts}}.
\bjtitle{Proceedings of the Astronomical Society of Australia}
\bvolume{4},
\bfpage{382}.
\doiurl{10.1017/S1323358000021214}.
\adsurl{https://ui.adsabs.harvard.edu/abs/1982PASAu...4..382S}.
\end{barticle}
\endbibitem

\bibitem[\protect\citeauthoryear{{The SunPy Community}
  \textit{et~al.}}{2020}]{sunpy_community2020}
\begin{barticle}
\bauthor{\bsnm{{The SunPy Community}}},
\bauthor{\bsnm{Barnes}, \binits{W.T.}},
\bauthor{\bsnm{Bobra}, \binits{M.G.}},
\bauthor{\bsnm{Christe}, \binits{S.D.}},
\bauthor{\bsnm{Freij}, \binits{N.}},
\bauthor{\bsnm{Hayes}, \binits{L.A.}},
\bauthor{\bsnm{Ireland}, \binits{J.}},
\bauthor{\bsnm{Mumford}, \binits{S.}},
\bauthor{\bsnm{Perez-Suarez}, \binits{D.}},
\bauthor{\bsnm{Ryan}, \binits{D.F.}},
\bauthor{\bsnm{Shih}, \binits{A.Y.}},
\bauthor{\bsnm{Chanda}, \binits{P.}},
\bauthor{\bsnm{Glogowski}, \binits{K.}},
\bauthor{\bsnm{Hewett}, \binits{R.}},
\bauthor{\bsnm{Hughitt}, \binits{V.K.}},
\bauthor{\bsnm{Hill}, \binits{A.}},
\bauthor{\bsnm{Hiware}, \binits{K.}},
\bauthor{\bsnm{Inglis}, \binits{A.}},
\bauthor{\bsnm{Kirk}, \binits{M.S.F.}},
\bauthor{\bsnm{Konge}, \binits{S.}},
\bauthor{\bsnm{Mason}, \binits{J.P.}},
\bauthor{\bsnm{Maloney}, \binits{S.A.}},
\bauthor{\bsnm{Murray}, \binits{S.A.}},
\bauthor{\bsnm{Panda}, \binits{A.}},
\bauthor{\bsnm{Park}, \binits{J.}},
\bauthor{\bsnm{Pereira}, \binits{T.M.D.}},
\bauthor{\bsnm{Reardon}, \binits{K.}},
\bauthor{\bsnm{Savage}, \binits{S.}},
\bauthor{\bsnm{Sipőcz}, \binits{B.M.}},
\bauthor{\bsnm{Stansby}, \binits{D.}},
\bauthor{\bsnm{Jain}, \binits{Y.}},
\bauthor{\bsnm{Taylor}, \binits{G.}},
\bauthor{\bsnm{Yadav}, \binits{T.}},
\bauthor{\bsnm{Rajul}},
\bauthor{\bsnm{Dang}, \binits{T.K.}}:
\byear{2020},
\batitle{The sunpy project: Open source development and status of the version
  1.0 core package}.
\bjtitle{The Astrophysical Journal}
\bvolume{890}.
\doiurl{10.3847/1538-4357/ab4f7a}.
\burl{https://iopscience.iop.org/article/10.3847/1538-4357/ab4f7a}.
\end{barticle}
\endbibitem

\bibitem[\protect\citeauthoryear{{Wild}}{1950}]{Wild1950}
\begin{barticle}
\bauthor{\bsnm{{Wild}}, \binits{J.P.}}:
\byear{1950},
\batitle{{Observations of the Spectrum of High-Intensity Solar Radiation at
  Metre Wavelengths. III. Isolated Bursts}}.
\bjtitle{Australian Journal of Scientific Research A Physical Sciences}
\bvolume{3},
\bfpage{541}.
\doiurl{10.1071/PH500541}.
\adsurl{https://ui.adsabs.harvard.edu/abs/1950AuSRA...3..541W}.
\end{barticle}
\endbibitem

\bibitem[\protect\citeauthoryear{{Zheleznyakov} and {Zaitsev}}{1970}]{Zhe1970}
\begin{barticle}
\bauthor{\bsnm{{Zheleznyakov}}, \binits{V.V.}},
\bauthor{\bsnm{{Zaitsev}}, \binits{V.V.}}:
\byear{1970},
\batitle{{Contribution to the Theory of Type III Solar Radio Bursts. I.}}
\bjtitle{\sovast}
\bvolume{14},
\bfpage{47}.
\adsurl{https://ui.adsabs.harvard.edu/abs/1970SvA....14...47Z}.
\end{barticle}
\endbibitem

\bibitem[\protect\citeauthoryear{{Zucca} \textit{et~al.}}{2012}]{Zucca2012}
\begin{barticle}
\bauthor{\bsnm{{Zucca}}, \binits{P.}},
\bauthor{\bsnm{{Carley}}, \binits{E.P.}},
\bauthor{\bsnm{{McCauley}}, \binits{J.}},
\bauthor{\bsnm{{Gallagher}}, \binits{P.T.}},
\bauthor{\bsnm{{Monstein}}, \binits{C.}},
\bauthor{\bsnm{{McAteer}}, \binits{R.T.J.}}:
\byear{2012},
\batitle{{Observations of Low Frequency Solar Radio Bursts from the Rosse
  Solar-Terrestrial Observatory}}.
\bjtitle{\solphys}
\bvolume{280}(\bissue{2}),
\bfpage{591}.
\doiurl{10.1007/s11207-012-9992-x}.
\adsurl{https://ui.adsabs.harvard.edu/abs/2012SoPh..280..591Z}.
\end{barticle}
\endbibitem

\end{thebibliography}
\bibliographystyle{spr-mp-sola}


\end{article}
\end{document}